\documentclass[twocolumn]{aastex63}
\usepackage{amsmath}
\usepackage{graphicx}
\usepackage{natbib}
\usepackage{txfonts}
\usepackage[referable]{threeparttablex}
\usepackage{tensor}
\usepackage{mathtools}
\usepackage{mathrsfs}
\usepackage{multirow}

\newcommand{\D}[0]{\mathrm{D}}

\renewcommand{\d}[0]{\mathrm{d}}

\renewcommand{\vec}[1]{\mathbf{#1}}

\begin{document}
\author[0000-0002-6917-0214]{K.~S.~Croker}
\author[0000-0001-8818-8922]{K.~A.~Nishimura}
\affiliation{Department of Physics and Astronomy, University of Hawai`i at M\=anoa,  2505 Correa Rd., Honolulu, Hawai`i, 96822 USA}
\author[0000-0003-1748-2010]{D.~Farrah}
\affiliation{Department of Physics and Astronomy, University of Hawai`i at M\=anoa,  2505 Correa Rd., Honolulu, Hawai`i, 96822 USA}
\affiliation{Institute for Astronomy, University of Hawai`i,  2680 Woodlawn Dr., Honolulu, Hawai`i, 96822 USA}

\correspondingauthor{K.~S.~Croker}
\email{kcroker@phys.hawaii.edu}

\shorttitle{The GEODE mass function}
\shortauthors{Croker, Nishimura, and Farrah}

\title{Implications of Symmetry and Pressure in Friedmann Cosmology. II. Stellar Remnant Black Hole Mass Function}

\begin{abstract}
We consider some observational consequences of replacing all black holes (BHs) with a class of non-singular solutions that mimic BHs but with Dark Energy (DE) interiors; GEneric Objects of Dark Energy (GEODEs).
We focus on the BH mass function and chirp-mass redshift distribution of mergers visible to gravitational wave observatories.
We incorporate the GEODE blueshift into an initially Salpeter stellar remnant distribution, and model the binary population by evolving synthesized binary remnant distributions, published before LIGO's first measurements.
We find that a GEODE produced between $20 \lesssim z \lesssim 40$, and observed at $z \sim 7$, will have its initial mass amplified by $\sim 20-140\times$.
This can relieve tension between accretion-only growth models and the inferred masses of BHs in quasars at $z \gtrsim 6$.
Moreover, we find that merger rates of GEODE binaries increase by a factor of $\sim 2\times$ relative to classical BHs.
The resulting GEODE mass function is consistent with the most recent LIGO constraints at $< 0.5\sigma$.
In contrast, a Salpeter stellar distribution that evolves into classical remnants is in tension at $\gtrsim 2\sigma$.
This agreement occurs without low-metallicity regions, abnormally massive progenitor stars, novel formation channels, or primordial object formation at extreme rates.
In particular, we find that solar metallicity progenitors, which produce $1.1-1.8\mathrm{M}_\odot$ remnants, overlap with many LIGO observations when evolved as GEODEs.
\end{abstract}


\section{Introduction}
\label{sec:intro}
Advanced LIGO and other planned observatories \citep[e.g.][]{dwyer2015gravitational} are providing an unprecedented census of compact objects.
Information on the population of black hole (BH) binaries is expected to inform models of core-collapse supernovae, binary system formation, and accretion/feedback mechanisms during subsequent evolution.
In the typical scenario \citep[e.g.][]{dominik2012double}, a progenitor binary stellar distribution at some redshift $z$ is processed via common envelope and core-collapse physics into a remnant distribution.
In this way, tomographic information on the black hole mass function (BHMF) can be used to constrain the formation and growth of compact remnants.
The usual assumption is that the remnant objects are traditional BHs, which do not intrinsically evolve in time.

There are, however, alternative models to traditional black holes.
One such family of models are the GEneric Object of Dark Energy (GEODE) models, in which black hole interiors are described by a dark energy equation of state.
The simplest example of a GEODE is the de-Sitter sphere, first proposed by \citet{gliner1966algebraic} as a non-singular end stage of stellar gravitational collapse.
Another important example of a GEODE solution is the gravastar; formed in the limit of a radially decreasing sequence of Schwarzschild constant-density spheres \citep{mazur2015surface}.
\citet{posada2017slowly} has shown that this GEODE provides the only known viable source to the Kerr exterior spacetime, in the limit of slow rotation.
In general, thin-shell GEODEs can be stable to rotation \citep[e.g.][]{chirenti2008ergoregion, uchikata2015slowly, maggio2017exotic} and perturbation \citep[e.g.][]{visser2004stable, debenedictis2006gravastar, lobo2006stable}.
Finally, it is notable that GEODEs are not restricted to gravitational collapse, and can include ``vacuum bubbles,'' or isolated regions of energized vacuum \citep{berezin1987dynamics}.
These are examples of GEODEs, which need not even be stationary.  

Though observational signatures for GEODEs have been developed, they tend to be model dependent and degenerate.
With respect to the gravitational wave signals\footnote{\citet{collaborat2019first} have found an object with electromagnetic signatures consistent with the Kerr geometry.
    As pointed out by \citet[][Introduction]{chirenti2016did}, even the restricted class of stiff, thin-shelled GEODEs may not be distinguishable from black holes via electromagnetic signatures.} from merging objects, \citet{cardoso2016gravitational} caution that ringdown may not suffice to distinguish extremely compact remnants from classical BHs.
On the other hand, \citet{chirenti2016did} have argued that the ringdown of first observed merger, GW150914, is not consistent with any GEODE featuring a slowly-rotating, stiff shell.
See, however, \citet[][\S IV]{yunes2016theoretical}.
Subsequent investigations \citep[e.g.][]{cardoso2017tests} have explored possible echos after binary mergers, but \citet{PhysRevD.100.044027} demonstrate that echos may be absent.

An analysis of whether or not the details of ringdown or echos suffice to distinguish an arbitrary GEODE model from the classical black holes is beyond the scope of this paper.
We focus instead on a fundamental characteristic of GEODEs that leads to complementary and immediately testable consequences.
\citet[][\S4]{cw2018part1} establish that the members of any GEODE population, with non-vanishing averaged pressure, will cosmologically shift in energy.
In this paper we quantify and explore these consequences for such GEODEs with initial mass $\lesssim 100\mathrm{M}_\odot$.

In \S\ref{sec:blueshift}, we review the cosmological energy shift relevant for a GEODE.
We then present order of magnitude estimates to motivate our subsequent treatment and predict the apparent black hole mass function (BHMF) from first principles.
In \S\ref{sec:inspiral}, we derive the influence of local mass growth on bound Keplerian orbits involving one or two GEODEs.
In \S\ref{sec:geode_mergers}, we determine the observational consequences for gravitational wave observatories of both of these aspects of GEODE evolution.
In \S\ref{sec:outlook}, we briefly conclude. Appendices~\ref{sec:generic-bhmf} and \ref{sec:bhmf} develop a theoretical apparent BHMF, given an underlying GEODE population.
Appendix~\ref{sec:redshift-distribution} develops a distribution function in redshift for stellar production.
Appendix~\ref{sec:horizon} develops coarse LIGO detector selection criteria, as a function of chirp mass and merger redshift.
Appendix~\ref{sec:lowZ} studies low-metallicity progenitor binary GEODEs. Consistent with astrophysical literature, all densities and pressures will be comoving.
In other words, physical densities are multiplied by $a^3$ so that an ideal comoving dust density remains constant.
We often work in units where the speed of light $c \equiv 1$ and we have used the $(-, +, +, +)$ signature.
Since we reference it frequently, we abbreviate the first gravitational wave transient catalog produced by the \citet{ligo2018gwtc} as GWTC-1.

\section{The GEODE cosmological blueshift}
\label{sec:blueshift}
  Any GEODE population contributes to the cosmological energy density as
\begin{align}
   \rho_s(\eta) = \frac{1}{\mathcal{V}} \int_\mathcal{V} \sum_i \rho_i(\eta, \vec{x})~\d^3\vec{x}. \label{eqn:geode-contribution}
\end{align}
The above integral is expressed in RW coordinates with conformal time $\eta$, $\mathcal{V}$ is a suitably chosen cosmological volume, $i$ sums over all GEODEs enclosed in $\mathcal{V}$, and $\rho_i$ is zero outside of the $i$-th GEODE.
As shown by \citet[][Equation~(44)]{cw2018part1}, such a GEODE population also contributes an averaged pressure
\begin{align}
  \mathcal{P}_s(\eta) \equiv (-1 + \chi)\rho_s(\eta),
\end{align}
where $\chi$ encodes details of the particular GEODE model.

Given depletion of baryonic density $\Delta_\mathrm{b}(\eta)$ through stellar gravitational collapse to GEODEs, conservation of stress-energy at zero order in Friedmann cosmology takes the following form
\begin{align}
  \frac{\d \rho_s}{\d a} + 3(\chi - 1)\frac{\rho_s}{a} = \frac{\d \Delta_\mathrm{b}}{\d a} \label{eqn:model-scalefactor}.
\end{align}

The conservation of energy statement in Equation~(\ref{eqn:model-scalefactor}) can be immediately integrated for fixed $\chi$
\begin{align}
\rho_s = a^{3(1-\chi)}\int_0^a\frac{\d \Delta_\mathrm{b}}{\d a'}\frac{1}{a'^{3(1-\chi)}}~\d a'. \label{eqn:prediction}
\end{align}
Note that, as $\chi \to 1$, $\rho_s$ behaves as dust.
As $\chi \to 0$, however, the comoving density grows as $a^3$.
This is characteristic of DE: its physical density does not diminish with the expansion.

To develop some intuition for Equation~(\ref{eqn:prediction}) we consider a sequence of burst formations 
\begin{align}
\frac{\d \Delta_\mathrm{b}}{\d a} \equiv \sum_{n} Q_n \delta(a - a_n),
\end{align}
where $Q_n$ is the comoving density of baryons instantaneously converted at $a_n$.
Substitution into Equation~(\ref{eqn:prediction}) gives 
\begin{align}
  \rho_s(a) = \sum_n^{a_n \leqslant a} Q_n \left(\frac{a}{a_n}\right)^{3(1-\chi)}. \label{eqn:burst_density}
\end{align}
Explicitly, at the instant of the $m$-th conversion $a_m$, we have the following comoving density of DE 
\begin{align}
  \rho_s(a_m) = Q_m + \sum_n^{a_n < a_m} Q_n \left(\frac{a}{a_n}\right)^{3(1-\chi)}.
\end{align}
Thus, at the instant of conversion, $\rho_s$ agrees with a spatial average over the newly formed GEODEs' interiors.
In other words, a quantity of baryonic mass has been converted to an equal quantity of DE.
Beyond the instant of conversion, however, the contribution due to GEODEs dilutes more slowly than the physical volume expansion.

To understand this slowed dilution, we may write Equation~(\ref{eqn:geode-contribution}) for these burst productions
\begin{align}
\rho_s(a) = \frac{1}{\mathcal{V}} \int_\mathcal{V} \sum_n^{a_n \leqslant a} \sum_{i_n} \rho_{i_n}(a, \vec{x})~\d^3\vec{x}.
\end{align}
Here $i_n$ indexes over GEODEs produced in burst $n$.
Performing the spatial integral gives
\begin{align}
\rho_s(a) = \frac{1}{\mathcal{V}} \sum_n^{a_n \leqslant a} \sum_{i_n} M_{i_n}(a) \label{eqn:burst_geode_contribution},
\end{align}
where $M_{i_n}(a)$ is the mass of the $i$-th GEODE produced in burst $n$, at time $a$.
Noting that
\begin{align}
  Q_n = \frac{1}{\mathcal{V}} \sum_{i_n} M_{i_n}(a_n),
\end{align}
Equation~(\ref{eqn:burst_density}) becomes
\begin{align}
  \rho_s(a) = \frac{1}{\mathcal{V}} \sum_n^{a_n \leqslant a}  \sum_{i_n} M_{i_n}(a_n) \left(\frac{a}{a_n}\right)^{3(1-\chi)}.  \label{eqn:burst_density_mass}
\end{align}
Finally, equating Equation~(\ref{eqn:burst_geode_contribution}) and Equation~(\ref{eqn:burst_density_mass}) reveals that 
\begin{align}
  M_{i_n}(a) = M_{i_n}(a_n)\left(\frac{a}{a_n}\right)^{3(1-\chi)} \label{eqn:blueshift}.
\end{align}
If we set $\chi \to 0$, the mass grows as $a^3$.
In other words, an inertially propogating GEODE cosmologically blueshifts.

There is ample precedent for such general relativistic effects in RW cosmology: an inertially propogating photon cosmologically redshifts.
Note that the Strong Equivalence Principle \citep[e.g.][\S3.3]{TEGP} is not violated, because a quasilocal Lorentz frame cannot be defined on cosmological timescales.
There is precedent in GR for cosmologically coupled evolution within compact object models.
For example, \citet{nolan1993sources} constructs an $O(1)$ interior source which joins correctly to the arbitrary FRW-embedded point mass solution of \citet{McVittie1933}.
Its interior densities and pressures evolve cosmologically\footnote{In the case of Nolan's solution, they evolve so as to keep the mass of the object fixed.
  This follows from constraints imposed by McVittie in the construction of his original solution.
  His specific intent was to keep the mass fixed.}.

  In the scenario described by Equation~(\ref{eqn:model-scalefactor}), this shift occurs to all material in a RW cosmology described by GR, but it is only relevant to ultra-relativistic material.
For example, consider the core of a typical star.
By ideal gas arguments \citep[e.g.][\S4.1.1]{cw2018part1}, typical stellar material has an equation of state $w \equiv \chi - 1$ such that
\begin{align}
  1 \leqslant \chi \lesssim 1 + 10^{-7}.
\end{align}
This means that, between $z=1$ and today, the gravitating mass of a star will decrease, at most
\begin{align}
  M(0) = M(1)2^{-3\times10^{-7}} = M(1)\left(1 - 2\times 10^{-7}\right).
\end{align}
The change is minuscule.
This means the change for typical stars, planets, and gas is completely unobservable even on cosmological timescales.

\subsection{Order of magnitude estimates}
\label{sec:oom}
Before detailed analysis of the GEODE blueshift, we consider observable situations that can be quickly treated with order of magnitude calculations.
This will establish measurements that can confirm or exclude the GEODE blueshift.

\subsubsection{Low-redshift mass boundaries}
We first examine the effect of the GEODE blueshift on the low-redshift stellar remnant population.
To do so we consider a GEODE of mass $1.5\mathrm{M}_\odot$ produced at $z_i = 2$.
This mass is typical of the expected initial mass of remnants of high-mass stars \citep[e.g.][]{fryer2012compact} while the redshift is within the peak of the comoving star formation rate density.
If this GEODE is then observed at $z_\mathrm{obs} = 0.18$ then the mass arising solely from the GEODE blueshift will be
\begin{align}
  1.5\mathrm{M}_\odot \frac{(1 + 2)^3}{(1 + 0.18)^3} = 24.65\mathrm{M}_\odot. %2*12.167/1.6343
\end{align}
The GEODE blueshift will thus have a profound impact on the low redshift remnant population, decreasing the number of low mass objects.

For example, blueshift is expected to grow remnants to masses $m \in (50, 135)~\mathrm{M}_\odot$.
This is to be contrasted with results from stellar evolution and core collapse models that predict no such remnants.
In the GEODE scenario, however, objects in excess of $50\mathrm{M}_\odot$ may be present at low redshifts, as they can form from, e.g., a $3\mathrm{M}_\odot$ remnant produced at $z_i = 2$.
There is no need for multiple mergers within dense stellar clusters \citep[e.g.][and references therein]{mapelli2016massive} to populate the mass gap.
Whether such objects can be observed in mergers is a more subtle question, addressed at length in \S\ref{sec:inspiral}.  
In any case, the existence of large-mass remnants at low-redshift cannot be used directly to make inferences about core collapse physics or progenitor environments \citep[e.g.][]{PhysRevD.100.041301}.

\subsubsection{High-redshift quasars and supermassive black holes}
The observed masses of supermassive black holes (SMBHs) in quasars at high redshift is in tension with theoretical expectations.
The GEODE blueshift can help relieve these tensions.
Roughly 40 quasars at $z \geqslant 6$ with mass $> 10^9 \mathrm{M}_\odot$ have been observed \citep[e.g.][]{mortlock2011luminous, 2003ApJ...587L..15W, wu2015ultraluminous}.
As reviewed by \citet{volonteri2012formation}, the production of such objects poses significant challenges to theoretical models, requiring some combination of effects such as sustained super-Eddington accretion, and runaway collapse of nascent star clusters.
The GEODE cosmological blueshift will however contribute a factor of
\begin{align}
  \frac{(z_i + 1)^3}{(7 + 1)^3}
\end{align}
to the object's mass at $z_\mathrm{obs}=7$.
For a rare but still feasible $z_i=40$ object, this gives a mass amplification of $134\times$.
This is a lower bound, because accreted mass during $7 \leqslant z \leqslant z_i$ will also blueshift.
For GEODEs at $z_i=20$, the lower-bound mass amplification is a still-significant $18\times$.
Moreover, the existence of these rare objects does not meaningfully influence $\Omega_\Lambda$.
Assuming a number density of $0.3~\mathrm{Gpc}^{-3}$ at $z=6$ \citep{jiang2016final} and an average mass of $10^9~\mathrm{M}_\odot$, this becomes $0.3~\mathrm{M}_\odot\cdot\mathrm{Mpc}^{-3}$.
This compares negligibly to the critical density today of $1.5\times 10^{11}~\mathrm{M}_\odot\cdot\mathrm{Mpc}^{-3}$.

Assuming the same number density and average mass at $z=6$ as above, then, at $z<0.2$ (Gpc comoving), we expect to observe at least seven objects with masses of at least $2\times10^{11}\mathrm{M}_\odot$.
This lower bound comes from assuming growth soley via the GEODE blueshift.
This is a factor of at least two more massive than the most massive black hole currently known in the nearby universe, reported by \citet{dullo2017remarkably}.
Large-scale surveys of apparent BH mass demographics in the low redshift universe are thus a straightforward way to test the GEODE hypothesis.

The GEODE scenario also argues against accretion as the dominant growth mechanism for less extreme GEODEs in galactic centers.
Consider a $10\mathrm{M}_\odot$ GEODE at $z_i = 20$ that grows soley via the GEODE blueshift.
The mass of this object at $z=0$ is 
\begin{align}
  10\mathrm{M}_\odot (1 + 20)^3 = 0.9\times 10^5~\mathrm{M}_\odot. %2*12.167/1.6343
\end{align}
This is within a factor of 20 of the current mass of the supermassive BH Sgr A*, and implies that mergers, accretion, and the GEODE blueshift all play a role in growing SMBHs in all types of galaxy.

\subsection{Theoretical GEODE mass function given a Salpeter progenitor distribution}
\label{sec:thenut}
\begin{figure}[t]
\centering
\includegraphics[width=\linewidth]{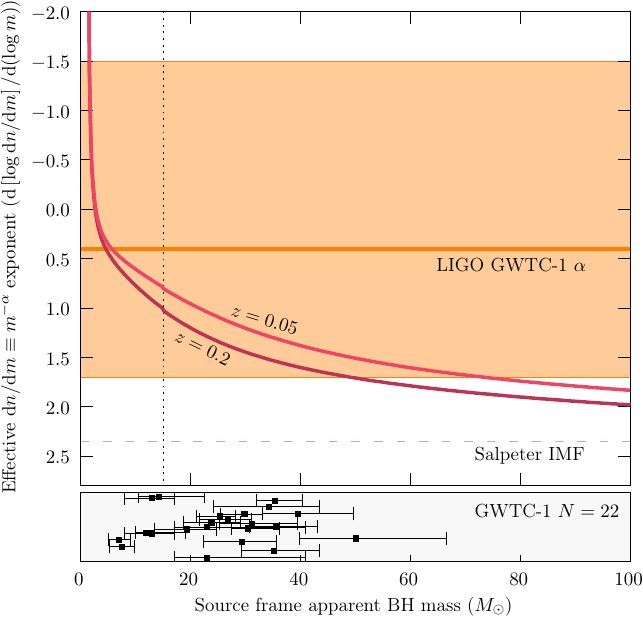}
\caption{\label{fig:bhmf}Slope of the $\log$-$\log$ plot of apparent black hole mass functions (BHMFs), given a $\chi = 0$ GEODE population.
  In this plot, pure power-laws appear as horizontal lines and the vertical axis is inverted
  Predicted BHMFs assume an initial remnant population, which directly tracks a constant Salpeter $\alpha = 2.35$ progenitor distribution (thin dashed).
  Upper mass cap for the initial remnant distribution is $m_h = 15\mathrm{M}_\odot$ (thin dotted).
  The predicted BHMFs evolve in redshift and are labeled accordingly.
  These predictions neglect mergers and accretion, so the indicated curves give upper bounds for the effective power-law $\alpha$.
  The GWTC-1 sample of $N=22$ primary/secondary binary pair masses is indicated beneath the plot, with uncertainties at $90\%$ confidence.
  All GWTC-1 BH-BH mergers occurred within or near the indicated redshifts.
  The LIGO best fit power-law to this sample is $\alpha = 0.4^{+1.3}_{-1.9}$ indicated in orange (dark gray) at $90\%$.
  It is in tension with a Salpeter distribution at $\gtrsim 2\sigma$.}
\end{figure}%2
The previous order of magnitude observations motivate construction of a theoretical model for the mass and redshift distribution of GEODEs.
We construct such a model in Appendices~\ref{sec:generic-bhmf} and \ref{sec:bhmf}.
This model assumes no mergers or accretion, and that the remnant formation rate tracks the mean density in stars $\rho_*$.
We assume that the initial remnant mass distribution tracks a progenitor Salpeter $m^{-\alpha}$ distribution, with $\alpha = 2.35$ constant in time.
This model, given in Equation~(\ref{eqn:bhmf}), is completely independent of any LIGO BH measurements.
Evaluated for a $\chi = 0$ GEODE population, we find
\begin{align}
  \frac{\d n}{\d m}&\propto m^{-\alpha} a^{3(\alpha - 1)} \int_{a(m_\ell/m)^{1/3}}^{a\min\left[1,~(m_h/m)^{1/3}\right]} \frac{\d \rho_*}{\d a_0} a_0^{3(1-\alpha)}~\d a_0, \label{eqn:bhmf_floating_normal}
\end{align}
where $m_\ell$ and $m_h$ are the lower and upper cutoff masses for the initial remnant distribution, respectively.
Distributions computed for $z=0.2$ and $z=0.05$ are displayed in Figure~\ref{fig:bhmf}, where we have set $m_\ell = 1.1\mathrm{M}_\odot$ and $m_h =15\mathrm{M}_\odot$ consistent with the range of remnants studied by \citet{de2015merger}.

The qualitative features described in \S\ref{sec:oom} are evident in Figure~\ref{fig:bhmf}, where the mass function grows more top heavy upon approach to the present epoch.  
Promising agreement with the measured $\alpha$ reported in GWTC-1, \S\phantom{}VIIB, is apparent.
Note that these curves are upper bounds on $\alpha$, since mergers and accretion have been neglected.
The jump discontinuity at $m_h = 15\mathrm{M}_\odot$ is due to the capped remnant distribution assumed in Equation~(\ref{eqn:capped_remnants}).
It reflects that the production of remnants below $m_h$ is a combination from an initial population and a blueshifted population with initially smaller mass.
Masses above $m_h$ can only result via blueshift.
The results are not particularly sensitive to $\alpha$ or $m_h$, but are somewhat sensitive to $m_\ell$, as expected from any power-law.

From even this very simple model, we may predict that
\begin{itemize}
\item{maximum likelihood fits for the $\alpha$ in Equation~(\ref{eqn:bhmf}) will regenerate $\alpha \sim 2.3$, inherited from the stellar IMF}
\item{the errors on these fits will be smaller than those for $\alpha$ fit against $m^{-\alpha}$ alone}
\item{the measured distribution evolves in redshift}
\end{itemize}
In summary, investigation of the mass function is definitive for identifying the GEODE scenario.

\section{GEODE blueshift inspiral in binaries}
\label{sec:inspiral}
The cosmological mass gain of GEODEs will significantly alter their Keplerian orbits within double compact object (DCO) systems.
In this section, we first show that the effect is an adiabatic inspiral.
We then show how to incorporate this effect into existing gravitational wave orbital decay equations, in the linearized gravity regime.

In any DCO involving a GEODE, the masses are changing in time, but the Lagrangian contains no explicit orbital angle $\phi$ dependence.
This guarantees that angular momentum $L$ remains conserved.
While energy is no longer conserved due to explicit time dependence in the mass, the change is always negligable over one orbital period.
It follows that the squared angular momentum is still given by the typical \citep[e.g.][Equations~(15.4), (15.6)]{LandauMechanics} expression
\begin{align}
  L^2 &= GR(1-e^2)\mu m_1m_2,
\end{align}
where $\mu$ is the reduced mass
\begin{align}
  \mu &\equiv \frac{m_1 m_2}{m_1 + m_2},
\end{align}
$G$ is the gravitational constant, $R$ is the semi-major axis of the elliptical orbit, and $e$ is the eccentricity.
We conclude that $R(1-e^2)$ must change inversely to $\mu m_1 m_2$ to keep the angular momentum constant.

To determine the relation between $R$ and $e$, we use the theory of adiabatic invariants.
\citet{sivardiere1988adiabatic} constructs two simple adiabatic invariants for the Kepler problem
\begin{align}
  J_\phi \equiv \frac{4\pi^2 \mu R b}{P} \qquad J \equiv 2|E|P, 
\end{align}
where $b$ is the semi-minor axis of the ellipse, $E$ is the energy, and $P$ the orbital period.
Their ratio is necessarily invariant
\begin{align}
  \frac{J_\phi}{J} = \frac{4\pi^2 R^3}{P^2}\frac{\mu}{2|E|} \frac{b}{R^2}.
\end{align}
Here we have multiplied and divided by $R^2$ and grouped for clarity.
Substitution of Kepler's third law gives
\begin{align}
  \frac{J_\phi}{J} = \frac{Gm_1 m_2}{2|E|} \frac{b}{R^2}.
\end{align}
Finally, because
\begin{align}
  R = \frac{Gm_1m_2}{2|E|},
\end{align}
we find that
\begin{align}
  \frac{J_\phi}{J} = \frac{b}{R}
\end{align}
is invariant.
We conclude that the shape of the ellipse, and therefore $e$, is unchanged during GEODE blueshift inspiral.
Solving for $R$ gives
\begin{align}
  R = \left(\frac{L^2}{G}\right)\frac{m_1 + m_2}{\left(m_1m_2\right)^2(1-e^2)}.
\end{align}

Let the primary mass $m_1$ and secondary $m_2 < m_1$ possibly blueshift according to Equation~(\ref{eqn:blueshift}).
We will approximate neutron stars and all non-GEODE objects as intrinsically fixed in mass\footnote{As discussed in \citet[][\S4.2]{cw2018part1}, NSs are expected to exhibit a very small loss in energy because their averaged equation of state $w \sim 0.07$.  This loss may be accessible to precision timing experiments.  At the precision we consider in this treatment, we may neglect this effect.}.
We will only consider pure de-Sitter GEODEs.
These two conditions are reflected as follows
\begin{align}
  \chi \simeq \begin{cases}
    1 & \mathrm{NS} \\
    0 & \mathrm{GEODE}
  \end{cases}. \label{eqn:eos_assumptions}
\end{align}
We then find the following behaviors for the semi-major axis
\begin{align}
  R(a) = \frac{L^2}{G(1-e^2)}\begin{dcases}
    \frac{m_1 + m_2}{(m_1 m_2)^2} & \text{NS-NS} \\
    \frac{m_1(a/a_1)^3 + m_2}{(m_1 m_2)^2} \left(\frac{a_1}{a}\right)^6  & \text{NS-GEODE} \\
    \frac{m_1 + m_2}{(m_1 m_2)^2}\left(\frac{a_1}{a}\right)^9 & \text{GEODE-GEODE} 
  \end{dcases}. \label{eqn:geode_inspirals}
\end{align}
Here $a_1$ is the formation time of the GEODE or GEODEs.
In the GEODE-GEODE case, we have regarded these times as the same for clarity.
There is no loss of generality because the GEODE's mass in an NS-GEODE system can be used as the initial mass of the GEODE-GEODE system 
\begin{align}
  m_2 \left(\frac{a}{a_2}\right)^3 = \left[m_2 \left(\frac{a_1}{a_2}\right)^3\right]\left(\frac{a}{a_1}\right)^3. 
\end{align}
In other words, one can just redefine the ``initial'' mass of $m_2$ to be its mass at the formation time for the GEODE with mass $m_1$ for $a \leqslant a_1$.

The orbital evolution of a DCO, under the influence of both GEODE blueshift and GW losses is now straightforward. 
Regard the semi-major axis as $R(L, e, m_1, m_2)$, so that
\begin{align}
  \frac{\d R}{\d a} = \underbrace{\left[\frac{\partial R}{\partial L}\frac{\partial L}{\partial a} + \frac{\partial R}{\partial e}\frac{\partial e}{\partial a} \right]}_{\text{Gravitational radiation}} + \underbrace{\left[\frac{\partial R}{\partial m_1}\frac{\partial m_1}{\partial a} + \frac{\partial R}{\partial m_2}\frac{\partial m_2}{\partial a}\right]}_{\text{GEODE blueshift}} \label{eqn:orbit-evolution}
\end{align}
Equations for orbital decay by GW emission in linearized gravity are given in elementary texts, such as \citet[][Equations~(9.167)--(9.168)]{schutz2009first}.
These can be converted to scale factor with $1/aH$.
Since these equations for $\d R/d a$ due to GW emission assume fixed $m_1$ and $m_2$, they can be used for the GW loss term immediately with substituted time-dependent masses.
The time-evolution of $L$ and $e$ is given entirely by the usual expressions, again with substituted time-dependent masses.
This follows because, under the GEODE blueshift, $L$ is conserved and $e$ is fixed.

\section{Population estimates for merging binaries involving a GEODE}
\label{sec:geode_mergers}
The orbital decay timescale for both NS-GEODE and GEODE-GEODE binaries due to blueshift is much shorter than that for GW wave orbital decay until very near to merger.
The effect is cosmological though, so it is never visible during the final inspiral, merger, or ringdown.
This is a significant effect, which has consequences for the population of coalescing NS-GEODE and GEODE-GEODE binaries observable by interferometers.
Qualitatively, we expect two effects.
First, the masses of GEODEs will be increased relative to expected BH remnant masses.
Second, the number of binary systems which are capable of merging is increased.
This is because GEODE binaries at initially larger separation $R_i$ will inspiral due to GEODE blueshift until they can merge via GW orbital decay.
Competing with this increase, however, is that inspiral can be rapid enough to deplete GEODE binaries.
In other words, GEODE binaries may now merge earlier than expected and thus lie outside of detector sensitivities in amplitude and frequency.
The predicted population of NS-NS mergers should agree with existing estimates, because we regard NS masses as intrinsically fixed.

To quantify the effect of blueshift on GEODE double compact objects (DCOs), we begin by reviewing the standard DCO formation scenario.
Comprehensive studies by \citet{dominik2012double, dominik2013double, dominik2015double} and \citet{de2015merger} investigate
\begin{enumerate}
\item{an initial distribution of stellar binaries in: mass, mass ratio, period, eccentricity, and metallicity}
\item{evolution of these binaries through a common envelope (CE) phase, which decays the orbit and allows mass transfer}
\item{eventual core collapse/supernova (SN) of both objects into BHs or NSs} 
\end{enumerate}
These studies naturally assume that the masses of the constituent compact objects do not intrinsically change.
In a GEODE scenario, following Equation~(\ref{eqn:eos_assumptions}), this assumption remains valid until one member of the pair has undergone core-collapse to a GEODE.

\subsection{Initial conditions from CE and SN models}
\begin{table}[t]
  \centering
  \begin{threeparttable}
    \caption{\label{tbl:dco-masses}Mass ranges within de Mink \& Belczynski $Z_\odot$ DCO initial conditions}
    \begin{tabular}{ccc}
      \toprule
      Initial condition & Primary $(\mathrm{M}_\odot)$ & Secondary $(\mathrm{M}_\odot)$ \\
      \hline
      BH-BH & $5.4\leqslant m_1 \leqslant 15$ &  $5.4\leqslant m_2 \leqslant 15$ \\
      BH-NS &  $5.7\leqslant m_1 \leqslant 12.8$ & $1.1\leqslant m_2 \leqslant 1.9$ \\
      NS-NS & $1.1\leqslant m_1 \leqslant 1.8$ & $1.1\leqslant m_2 \leqslant 1.6$
    \end{tabular}
    \tablecomments{Remnant masses $1.8\mathrm{M}_\odot < m < 5.4\mathrm{M}_\odot$ were excluded \emph{a priori} by restriction to a supernovae collapse engine which excludes these remnants.}
  \end{threeparttable}
\end{table}%
To facilitate comparison of the GEODE scenario with the typical BH scenario, we will use the simulated DCO populations of \citet{de2015merger} as initial conditions.
This approach has many advantages.
First, this data set contains a wealth of initial conditions over observationally well-motivated ranges of progenitor parameters.
Further, these progenitor distributions are processed through intricate astrophysics during the CE phase.
Since the resulting DCO distribution may be correlated in remnant parameters, the most immediate comparison to previous work is to use exactly their same DCO population.
Second, this data set uses pre-LIGO era assumptions based on established observations of stellar populations.
In other words, it is blinded to BH measurements except for these authors' use of X-ray binary (XRB) inferred BH masses to preferentially select a supernovae collapse model~\citep[i.e.][]{belczynski2012missing}. 
This enables us to determine the influence of the GEODE blueshift in models free from the assumption that LIGO's observed mergers must be classical BHs.
Third, this data set is completely available to the public and so our assertions can be verified and extended.

The individual progenitor initial condition data sets are detailed in \citet[][\S2.1]{de2015merger}.
We use exclusively the N class models because their results are mostly insensitive to investigated variations in these models.
We use all three model A data sets: NS-NS, BH-NS, and BH-BH.
These labels indicate the possible mass ranges of remnants, and are displayed in Table~\ref{tbl:dco-masses}.
Model A includes DCOs that initiate mass transfer during departure from the main sequence (i.e. crossing the Hertzsprung gap).
While this almost certainly occurs, there is very little observational constraint within the Hertzsprung gap, because large stars cross this gap very rapidly.
To accommodate this uncertainty and develop lower-bound estimates on merger rate, \citet{de2015merger}  consider a distinct Model B that excludes these systems.
We do not consider model B in this treatment.

In order to produce a GEODE during core-collapse, core material must be converted into Dark Energy (DE).
Since this process is completely unconstrained at present, the mass boundary between neutron star and GEODE remnants is unclear.
In the simulations of \citet{de2015merger}, labels were applied strictly based on final remnant mass.
Similarly, LIGO observations at present can only distinguish NSs from BHs/GEODEs via their chirp frequency.
In order to study the signatures of a lower-mass GEODE population, we consider all initial conditions as $N_\mathrm{GEODE} = 1$ and $N_\mathrm{GEODE} = 2$ binaries.  
\begin{table}[t]
  \centering
  \begin{threeparttable}
    \caption{\label{tbl:merger-fractions}GEODE merger fractions relative to control de Mink \& Belczynski $Z_\odot$ DCO populations}
    \begin{tabular}{cccc}
      \toprule
      &&\multicolumn{2}{c}{Merged fraction} \\
      Initial condition & $N_\mathrm{GEODE}$ & (actual) & (observable)\\
      \hline
      BH-BH & 2 & $3.91$ & $211$ \\
      BH-NS & 2 & $2.99$ & $118$\\
      NS-NS & 2 & $1.64$ & $36.7$\\
      \hline
      BH-NS & 1 & $1.16$ & $0.206$ \\
      NS-NS & 1 & $1.48$ & $0.910$
    \end{tabular}
    \tablecomments{The BH-NS initial conditions, with $N_\mathrm{GEODE}=1$ has $N=7$ events after selection. BH-NS with fixed cosmological mass has $N = 34$ events after selection.  The rate (observed) for BH-NS thus has large uncertainties.}
  \end{threeparttable}
\end{table}%
For objects that do not intrinsically gain mass, their creation time does not influence the resulting population apart from metallicity effects.
This motivates the typical time-uniform assumption.
The GEODE spectrum, however, depends on when the GEODEs are formed.
Due to this time dependence, the SFR may provide a more accurate representation of the underlying population.
For each set of initial conditions, we sample in redshift from a distribution built from the cosmic star formation rate as given by \citet[][Equation~(15)]{MadauDickinson2014}.
This distribution is constructed in Appendix~\ref{sec:redshift-distribution}.
A time-uniform redshift distribution leads to few qualitative changes, apart from a preponderance of very low $z < 0.02$ mergers in GEODE and control scenarios.

We focus on solar metallicity ($Z_\odot$) initial conditions, since such progenitors are expected to dominate the stellar population.  
Results for low-metallicity ($0.1Z_\odot$) initial conditions are given in Appendix~\ref{sec:lowZ}, where we verify the known consistency with control populations and the LIGO sample in mass, chirp-mass, and redshift. 

\subsection{Methods}
We evolve the DCO initial conditions in redshift according to Equation~(\ref{eqn:orbit-evolution}).
These initial conditions have primary and secondary remnant masses with ranges given in Table~\ref{tbl:dco-masses}.
Our control population is built from the same DCO initial conditions, with masses held fixed.
Each initial condition is evaluated at 100 distinct redshifts sampled from Equation~(\ref{eqn:redshift-distribution}).
We limit redshift samples to $z_i < 4$ because GEODEs formed at high redshift merge at redshifts beyond current generation detector sensitivities.
In simulations with GEODEs, we always regard the more massive object, at DCO system formation, as a GEODE.
We neglect the very short time, relative to cosmological scales, between formation of the first and second compact object.
We use Python 3.5.3 and integrate via \texttt{odeint}\footnote{all simulation, analysis, and visualization code is available at \url{http://www.github.com/kcroker/cnf-geodes} under GPL v3 licensing.  We also include reference data sets, available elsewhere, for convenience.}.
We terminate each evolution, via a raised exception, when its periastron has dropped below $10^{-2}~\mathrm{AU}$ (merger) or when $z < 10^{-3}$ (no merger), whichever occurs first.
We then apply selection criteria to the merging population based on the LIGO O1 luminosity distance horizon, converted to redshift and adjusted for unequal primary and secondary masses.
These selection criteria are described in Appendix~\ref{sec:horizon}.

\subsection{Results}
\begin{table}[t]
  \centering
  \begin{threeparttable}
    \caption{\label{tbl:merger-rates}Predicted rates of $Z_\odot$ progenitor remnants vs. LIGO GWTC-1 rates (90\% confidence)}
    \begin{tabular}{ccccc}
      \toprule
      & Initial & & \multicolumn{2}{c}{Merger rate $(\mathrm{Gpc}^{-3}~\mathrm{yr}^{-1})$}  \\
      &  condition & $N_\mathrm{GEODE}$ & \phantom{ZZ}Sample\phantom{ZZ} & GWTC-1 \\
      \hline
      & BH-BH & $2$ & $109$ & $110-3840$\\
      & BH-NS & $2$ & $8.37$ & $110-3840$\\
      & NS-NS & $2$ & $100$ & $110-3840$ \\
      \hline
      & BH-NS & $1$ & $3.25$ & $< 610$\\
      & NS-NS & $1$ & $90.3$ & $< 610$ \\
      \hline
      \multirow{3}{*}{\rotatebox{90}{(Control)}} & BH-BH & $0$ & $28$ & $110-3840$\\
       & BH-NS & $0$ & $2.8$ & $< 610$ \\
       & NS-NS & $0$ & $61$ & $9.7-101$
    \end{tabular}
  \end{threeparttable}
\end{table}%
In this section we report the results of processing the DCO population through Keplerian orbital decay.
First, we will examine merger fractions and rates relative to the control population.
This will allow us to verify our qualitative expectations of the GEODE scenario.
Then we will present the BH-BH, BH-NS, and NS-NS initial conditions, evolved with $N_\mathrm{GEODE} = 2$, in detail.
We exclude the BH-NS population, evolved with $N_\mathrm{GEODE}=1$, from detailed consideration.
It has very low statistics $(N=7)$ due to the development of large mass ratios.

\subsubsection{Merger fractions and rates}
\label{sec:fractions_and_rates}
\begin{figure}[t]
  \centering
  \includegraphics[width=\linewidth]{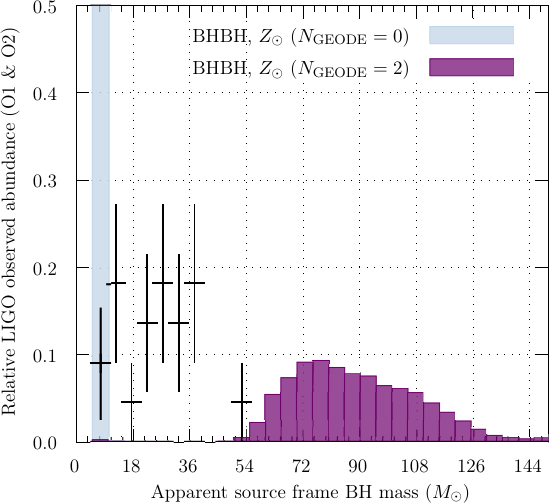}
  \caption{\label{fig:bhbh-geodegeode} Relative abundance vs. source frame mass for the BH-BH initial conditions of \citet{de2015merger}, evolved to merger assuming $N_\mathrm{GEODE}=2$.
    GEODE population is shown in purple (dark grey), control population evolved with cosmologically fixed mass is shown in blue (light grey).
    LIGO/Virgo GWTC-1 catalog overlaid with Poisson errors (black crosses).
    Populations have been selected according to LIGO O1 detector sensitivities following the procedure in Appendix~\ref{sec:horizon}.
    Graph is truncated at $50\%$ abundance for clarity, the control population bin extends to $100\%$.}
\end{figure}%
\begin{figure}[t]
  \centering
  \includegraphics[width=\linewidth]{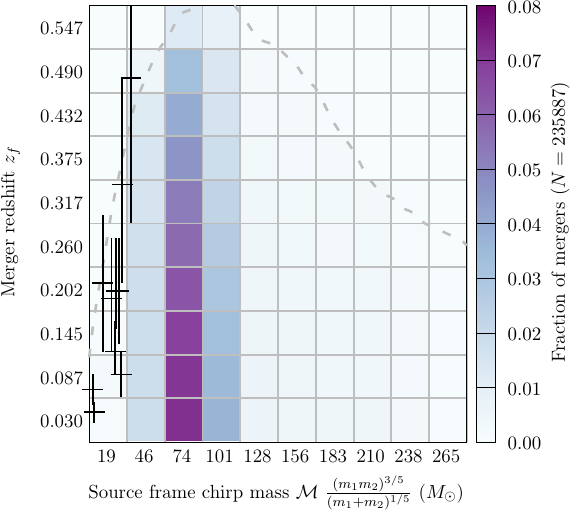}
  \caption{\label{fig:chirpz-bhbh-geodegeode} Merger redshift vs. source frame chirp mass for the BH-BH initial conditions of \citet{de2015merger}, evolved to merger assuming $N_\mathrm{GEODE}=2$.
    Figure axes represent all non-zero bins.
    LIGO/Virgo GWTC-1 catalog overlaid with 90\% credible intervals (black crosses).
    Fraction of mergers is encoded by color (intensity), the total number of events binned is given in the colorbar label.
    Populations have been selected (dashed line) according to LIGO O1 detector sensitivities following the procedure in Appendix~\ref{sec:horizon}.}
\end{figure}%
\begin{figure}[t]
  \centering
  \includegraphics[width=\linewidth]{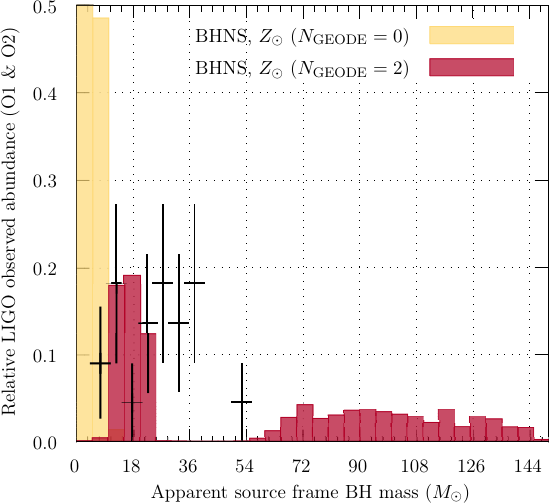}
  \caption{\label{fig:bhns-geodegeode} Relative abundance vs. source frame mass for the BH-NS initial conditions of \citet{de2015merger}, evolved to merger assuming $N_\mathrm{GEODE}=2$.
    GEODE population is shown in red (dark grey), control population evolved with cosmologically fixed mass is shown in orange (light grey).
    LIGO/Virgo GWTC-1 catalog overlaid with Poisson errors (black crosses).
    Populations have been selected according to LIGO O1 detector sensitivities following the procedure in Appendix~\ref{sec:horizon}.
    Graph is truncated at $50\%$ abundance for clarity, the control population bin extends to $100\%$.}
\end{figure}%
\begin{figure}[t]
  \centering
  \includegraphics[width=\linewidth]{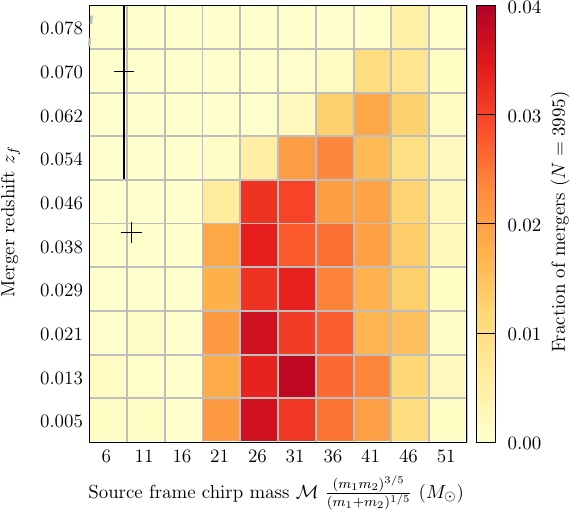}
  \caption{\label{fig:chirpz-bhns-geodegeode} Merger redshift vs. source frame chirp mass for the BH-NS initial conditions of \citet{de2015merger}, evolved to merger assuming $N_\mathrm{GEODE}=2$.
    Figure axes represent all non-zero bins.
    LIGO/Virgo GWTC-1 catalog overlaid with 90\% credible intervals (black crosses).
    Fraction of mergers is encoded by color (intensity), the total number of events binned is given in the colorbar label.
    Populations have been selected (dashed line) according to LIGO O1 detector sensitivities following the procedure in Appendix~\ref{sec:horizon}.}
\end{figure}%
\begin{figure}[t]
  \centering
  \includegraphics[width=\linewidth]{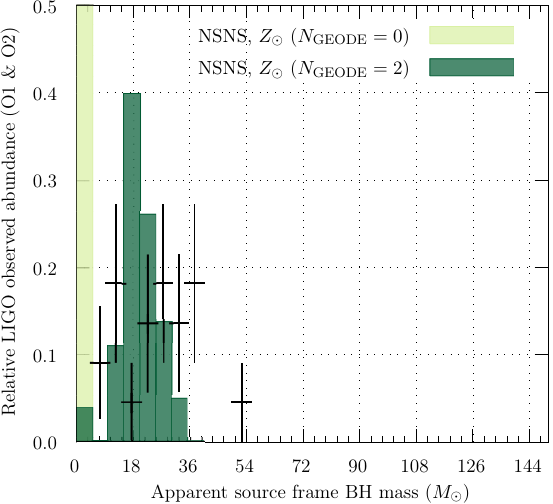}
  \caption{\label{fig:nsns-geodegeode} Relative abundance vs. source frame mass for the NS-NS binary population of \citet{de2015merger}, evolved to merger assuming $N_\mathrm{GEODE}=2$.
    GEODE population is shown in dark green (dark grey), control population evolved with cosmologically fixed mass is shown in yellow (light grey).
    LIGO/Virgo GWTC-1 catalog overlaid with Poisson errors (black crosses).
    Populations have been selected according to LIGO O1 detector sensitivities following the procedure in Appendix~\ref{sec:horizon}.
    Graph is truncated at $50\%$ abundance for clarity, the control population bin extends to $100\%$.}
\end{figure}%
\begin{figure}[t]
  \centering
  \includegraphics[width=\linewidth]{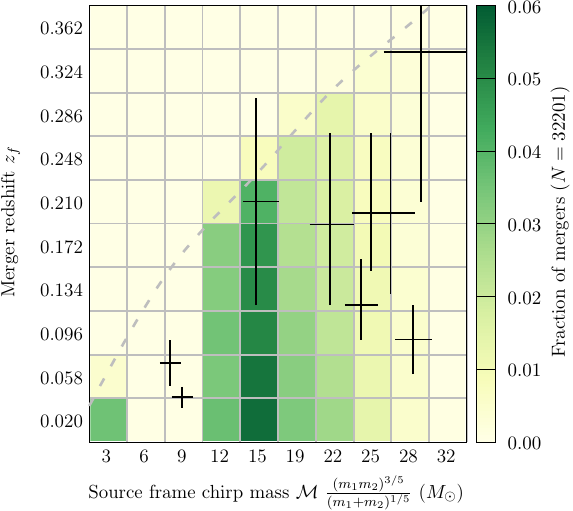}
  \caption{\label{fig:chirpz-nsns-geodegeode} Merger redshift vs. source frame chirp mass for the NS-NS binary population of \citet{de2015merger}, evolved to merger assuming $N_\mathrm{GEODE}=2$.
    Figure axes represent all non-zero bins.
    LIGO/Virgo GWTC-1 catalog overlaid with 90\% credible intervals (black crosses).
    Fraction of mergers is encoded by color (intensity), the total number of events binned is given in the colorbar label.
    Populations have been selected (dashed line) according to LIGO O1 detector sensitivities following the procedure in Appendix~\ref{sec:horizon}.}
\end{figure}%
The merging fraction of binary initial conditions evolved as GEODEs, relative to initially identical systems without cosmological blueshift, are displayed in Table~\ref{tbl:merger-fractions}.
In all cases, the number of actual mergers increases relative to control.
These increases are greatest when both objects are GEODEs.
This is expected, since GEODE blueshift inspiral allows capture of objects that could not merge by GW emission alone.
Note that $N_\mathrm{GEODE} = 1$ systems are less efficient at capture than $N_\mathrm{GEODE} = 2$ systems, as expected from Equation~(\ref{eqn:geode_inspirals}).

Observable mergers increase significantly in binary $N_\mathrm{GEODE} = 2$ settings.
This is because both objects have significantly larger masses at merger, and so their chirp mass lies within a more sensitive region of the detector. 
Observable mergers decrease in binary systems with only $N_\mathrm{GEODE} = 1$.
In this case, even though the chirp mass increases, a large mass ratio develops because only one object blueshifts.
This exaggerated ratio defeats the sensitivity improvement from higher chirp mass, as can be seen in Figure~\ref{fig:aligo_O1_sensitivity} and Equation~(\ref{eqn:massratio-effect}).

In Table~\ref{tbl:merger-rates}, we present merger rate estimates of binary systems containing one or two GEODEs.
We also include LIGO GWTC-1 merger rate measurements for comparison.
Actual merger rates are determined by scaling \citet[][Table~1, $\mathcal{R}(0)$]{dominik2015double} standard $Z_\odot$ metallicity rates by our computed merger fractions relative to control in Table~\ref{tbl:merger-fractions}.
We restrict comparison to these coarse estimates of \citet{dominik2015double} in order to capture the essential relative features of BHs vs. GEODEs within coalescing binary systems.

To put this table in context, the standard $Z_\odot$ NS-NS merger rate predicted by \citet{dominik2015double} is $61~\mathrm{Gpc}^{-3}~\mathrm{yr}^{-1}$.
This is in agreement with the GWTC-1 value of $9.7-101~\mathrm{Gpc}^{-3}~\mathrm{yr}^{-1}$ and suggests that the analysis of \citet{dominik2015double} accurately reflects this population.
The control standard $Z_\odot$ BH-BH merger rate, however, does not agree with the GWTC-1 90\% range.
For $N_\mathrm{GEODE}=2$ binary systems, BH-BH and NS-NS initial conditions nearly enter the GWTC-1 90\% range.
For $N_\mathrm{GEODE}=1$ binary systems, BH-BH and BH-NS initial conditions remain consistent with the GWTC-1 upper bound.

Note that our simplified merger criterion increases the merger rates for lower mass objects.
  This is because such systems radiate less via GW emission and take longer to inspiral below $10^{-2}~\mathrm{AU}$.
  To quanitify the effect of our approximation, we compared our merger rates for $N_\mathrm{GEODE} = 0$ (i.e. control) to those of \citet{de2015merger}.
  For the NS-NS populations, with $N_\mathrm{GEODE} = 0$, the rate increase was $0.7\%$.
  For the BH-BH populations, with $N_\mathrm{GEODE} = 0$, the rate increase was $0.3\%$.
  This confirms the expected qualitative behaviours and establishes the effect as subdominant.

\subsubsection{Observable GEODE populations vs. LIGO observations}
\begin{figure}[t]
  \centering
  \includegraphics[width=\linewidth]{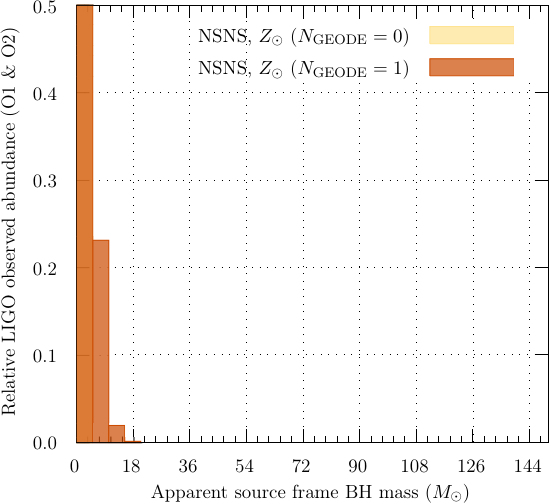}
  \caption{\label{fig:nsns-geodens} Relative abundance vs. source frame mass for the NS-NS binary population of \citet{de2015merger}, evolved to merger assuming $N_\mathrm{GEODE}=1$.
    GEODE population is shown in dark orange (dark grey), control population evolved with cosmologically fixed mass is shown in light orange (light grey).
    Populations have been selected according to LIGO O1 detector sensitivities following the procedure in Appendix~\ref{sec:horizon}.
    Graph is truncated at $50\%$ abundance for clarity, the control population bin extends to $100\%$.
    Note that LIGO has not yet observed any BH-NS mergers.}
\end{figure}

The observable populations of individual objects in merging binaries, built from the $Z_\odot$ metallicity progenitor initial conditions, are displayed in Figures~\ref{fig:bhbh-geodegeode}, \ref{fig:bhns-geodegeode}, \ref{fig:nsns-geodegeode}, and \ref{fig:nsns-geodens}.
Each of these histograms displays a control population (lighter color) with $N_\mathrm{GEODE} = 0$.
Abundance on the vertical axis has been cropped at $0.5$ for visibility.
Overlaid as black crosses on these figures are the binned GWTC-1 primary and secondary BH masses, with Poisson errors.

\begin{figure}[t]
  \centering
  \includegraphics[width=\linewidth]{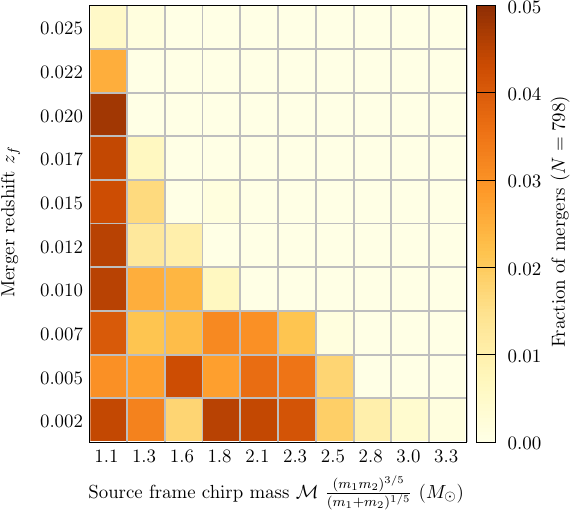}
  \caption{\label{fig:chirpz-nsns-geodens} Merger redshift vs. source frame chirp mass for the NS-NS binary population of \citet{de2015merger}, evolved to merger assuming $N_\mathrm{GEODE}=1$. 
    Relative density is normalized and encoded by color (intensity).  
    Populations have been selected according to LIGO O1 detector sensitivities following the procedure in Appendix~\ref{sec:horizon}.
    Note that LIGO has not yet observed any BH-NS mergers.}
\end{figure}

The observable populations of mergers, built from the $Z_\odot$ metallicity progenitor initial conditions, are displayed in Figures~\ref{fig:chirpz-bhbh-geodegeode}, \ref{fig:chirpz-bhns-geodegeode}, \ref{fig:chirpz-nsns-geodegeode}, and \ref{fig:chirpz-nsns-geodens}.
Each of these heatmaps displays the density of mergers in redshift and chirp mass bins, under the $N_\mathrm{GEODE} > 0$ scenario presented in the paired histogram.
The total number of simulated merger events is given in the colorbar axis label.
Heatmaps are identically zero off of the indicated bin ranges.
Overlaid as black crosses on these figures are the $N=11$ GWTC-1 BH-BH mergers, with $90\%$ confidence.
The selection criteria developed in Appendix~\ref{sec:horizon} is indicated by the dashed line, if visible.
Analysis of the raw merging sample reveals that the observed mass distribution is not sensitive to selection, apart from a proliferation of extremely low-mass binaries.

Figures~\ref{fig:bhbh-geodegeode} and \ref{fig:chirpz-bhbh-geodegeode} consider the BH-BH initial conditions, evolved as $N_\mathrm{GEODE}=2$ binaries.
Notice that the neither control nor GEODE populations agree with GWTC-1 data.
The control population is not massive enough, while the GEODE population is too massive.
In redshift-chirp space, there is rough agreement in merger redshifts, but the GEODE population is too massive.

Figures~\ref{fig:bhns-geodegeode} and \ref{fig:chirpz-bhns-geodegeode} consider the BH-NS initial conditions, evolved as $N_\mathrm{GEODE}=2$ binaries.
Again, the control population disagrees with observation.
The bimodal GEODE populations are clearly visible.
Notice that some GEODE masses begin to align with LIGO observations, but there is no agreement in chirp-redshift space.

Figures~\ref{fig:nsns-geodegeode} and \ref{fig:chirpz-nsns-geodegeode} consider the NS-NS initial conditions, evolved as $N_\mathrm{GEODE}=2$ binaries.
Again, the control population disagrees with observation.
This time, however, all GEODE masses are near to the LIGO observations but slightly too low.
This situation is confirmed in chirp-redshift space, where 9 of $N=11$ LIGO events are captured.
In redshift-chirp space, predicted merger redshifts and chirp masses are lower than the GWTC-1 events.

Figures \ref{fig:nsns-geodens} and \ref{fig:chirpz-nsns-geodens} consider the NS-NS initial conditions, evolved as $N_\mathrm{GEODE}=1$ binaries.
This configuration would represent apparent BH-NS binary mergers.
Since LIGO has not yet measured any such mergers, these figures represent predictions for this sub-population.

\subsection{Discussion}
\label{sec:discussion}
The most immediate observation is that the existing stellar population synthesis codes of \citet{dominik2012double, de2015merger} reproduce well the observed BH spectrum, if the remnant objects are initially lower mass GEODEs.
Specifically, the NS-NS data set at $Z_\odot$, evolved with $N_\mathrm{GEODE} = 2$, suggests that slightly heavier GEODE remnants in the mass range $1.8\mathrm{M}_\odot \leqslant m \leqslant 5.4\mathrm{M}_\odot$ represent a significant contribution to the GWTC-1 observed population.
This increased initial mass will bring individual GEODEs and GEODE binary merger chirps into the range bounded by the BH-BH and NS-NS initial conditions.
Further, more massive objects merge earlier in the $N_\mathrm{GEODE}=2$ scenario.
This increased redshift can be seen in the BH-BH initial conditions.
We predict
\begin{itemize}
  \item{initial conditions, which include the mass range $1.8\mathrm{M}_\odot \leqslant m \leqslant 5.4\mathrm{M}_\odot$, will give good agreement with the GWTC-1 observed population when evolved as $N_\mathrm{GEODE}=2$ binary systems.}
\end{itemize}
This range has typically been excluded based upon observations of remnant masses in X-ray binaries (XRBs).
The WATCHDOG sample from \citet{tetarenko2016watchdog}, however, contains 5 XRB systems with BH masses $< 5\mathrm{M}_\odot$.
This initial mass range is consistent with populations of such objects produced in the ``delayed'' SN simulations of \citet[][Figure~1]{belczynski2012missing}.

The next conclusion we may draw comes from Figures \ref{fig:bhbh-geodegeode} and \ref{fig:chirpz-bhbh-geodegeode}.
In these figures, initial remnant masses range from $5.4\mathrm{M}_\odot$ to $15\mathrm{M}_\odot$.
These initial remnant masses lead to observed masses at merger in excess of $50\mathrm{M}_\odot$.
In GWTC-1, there are no observed masses at merger beyond $55\mathrm{M}_\odot$, yet LIGO has good sensitivity in this range, as shown in Figure~\ref{fig:aligo_O1_sensitivity}.
From this, we may infer suppression of GEODE remnants with mass in excess $\sim$\phantom{}$5\mathrm{M}_\odot$.
Investigation of the merging population, before application of selection criteria, continues to support this position. 
A decreased maximum mass for remnants is consistent with lower-mass progenitor objects.
This increases abundances and removes the need to assume rare and large progenitors, which undergo pulsation instabilities~\citep[e.g.][]{belczynski2016effect}.

For the presumably young GEODEs in XRBs, blueshift will cause negligible deviations from the initial remnant mass.
Young objects with mass in excess of $5\mathrm{M}_\odot$, however, can still be produced through accretion.
The measured average mass transfer rates in the XRBs of the WATCHDOG sample\footnote{This catalog is available online in searchable and machine readable formats at \url{http://142.244.87.173}} range between $\sim 10^{-4}~\mathrm{M}_\odot~\mathrm{yr}^{-1}$ and $1~\mathrm{M}_\odot~\mathrm{yr}^{-1}$.
We suspect that the local XRB population provides an upward biased sample of remnant masses.

A remnant population with lower initial masses may also be consistent with state-of-the-art neutrino-neutron scattering calculations.
\citet[][Figure~7]{warrington2018} find that $E_\nu > 20~\mathrm{MeV}$ neutrino backscattering is suppressed by at least $10\times$, relative to previous estimates.
As is well known, modeling of the general relativistic magnetohydrodynamic collapse process is formidable.
The suppressed backscatter has not yet been incorporated into state-of-the-art SN collapse simulations \cite[e.g.][]{vartanyan2018successful}.
Decreased backscatter, however, is expected (C.~Fryer, \emph{priv. comm.}) to give a larger mass of stellar material sufficient velocity to escape accretion onto the compact remnant.
The final remnant will then have lower mass.

\section{Summary and outlook} 
\label{sec:outlook}

We have examined the observational consequences of replacing all BH stellar-collapse remnants with GEneric Objects of Dark Energy (GEODEs).
These ubiquitous solutions of General Relativity mimic the exteriors of BHs, but contain DE interiors.
Contrary to classical BHs, GEODEs may intrinsically gain mass via the same relativistic effect responsible for the photon redshift.
This blueshift, embodied in Equation~(\ref{eqn:blueshift}), supplements and amplifies any mass gained through typical accretion processes.

Our primary findings are as follows.
Population III GEODEs can grow by factors of $\sim$\phantom{}$100\times$ by redshift $z\sim 7$.
This can relieve tension between the observed masses of supermassive BHs in quasars at high redshift and their modeled formation timescales.
A simple GEODE mass function model at redshift $z < 4$, displayed in Figure~\ref{fig:bhmf}, agrees well with results from LIGO.
This model is built from initially Salpeter remnant distributions that blueshift proportional to the RW scale factor $a^3$.
Contrast this agreement with the anticipated Salpeter distribution for classical BHs, now in tension with observation at $\gtrsim 2\sigma$.
We predict that a maximum likelihood fit for model parameter $\alpha$ in Equation~(\ref{eqn:bhmf_floating_normal}) will reproduce the stellar IMF $\alpha\sim 2.3$.

The GEODE blueshift naturally produces the large masses observed in binary BH mergers.
Additionally, blueshift induces an adiabatic inspiral of Keplerian orbits that allows capture of wider binaries.
This effect amplifies merger rates by $\sim$\phantom{}$2\times$, relative to intrinsically fixed mass compact binaries.
The masses and merger redshifts of GEODE binaries are consistent with LIGO observations without low-metallicty, abnormally massive, or primordial progenitors.
These results are displayed in Figures~\ref{fig:nsns-geodegeode}~and~\ref{fig:chirpz-nsns-geodegeode}.
Current observations prefer an initial spectrum of lower-mass remnants without a gap after $1.8\mathrm{M}_\odot$, but suppressed above $5\mathrm{M}_\odot$.

A GEODE scenario may lead to observational consequences, which we have not considered in this treatment.
For example, \citet{mcconnell2013revisiting} have observed a shift in intercept of $M_\mathrm{BH}(\sigma)$ relations of early-type galaxies, relative to late-type galaxies.
In particular, early-type galaxies have a larger fraction of total mass within their central BHs.
This behavior is consistent with a GEODE scenario, because the older object will have blueshifted more.

\acknowledgments
K.~S.~Croker warmly acknowledges:
N.~Kaiser (ENS) for highlighting the possibility of adiabatic inspiral in GEODE systems;
J.~Barnes (IfA) for confirming quantitative aspects of adiabatic inspiral;
C.~Fryer (LANL) for discussions concerning remnant distributions;
C.~Pankow (Northwestern,~LIGO) for a beautiful summary of estimating LIGO detector sensitivities;
N.~Warrington (U.~Maryland) for clarifying the rapidly evolving investigations of core collapse and SN explosion;
G.~Tarle (U.~Michigan, DES) and A.~Romero-Wolf (JPL) for feedback on effective visualization;
J.~Kuhn (IfA) for comments on young binaries; and
E.~Avallone (IfA) and C.~Brinkman (IfA) for discussions about large stars.
This work was performed with financial support from the UH Vice Chancellor for Research.

\software{
  \href{https://www.scipy.org}{\texttt{scipy}}~\citep{scipy},
  \href{http://maxima.sourceforge.net}{Maxima}~\citep{maxima},
  \href{http://www.gnuplot.info}{gnuplot}~\citep{gnuplot}}

\appendix
\section{Generic BHMF model}
\label{sec:generic-bhmf}
To develop a model of the apparent BHMF, imagine a collection of GEODEs produced at an instant in time $a_0$.
The normalized distribution of these object's masses will be given by $f(m_0, a_0)$, such that
\begin{align}
  1 \equiv \int_0^\infty f(m_0, a_0)~\d m_0.
\end{align}
Since we are considering only a birth spurt, the total number of objects at any time $a$ is
\begin{align}
  N(a) = \int_0^\infty N(a_0)\Theta_{a_0}(a)f(m_0, a_0)~\d m_0.
\end{align}
Here $N(a_0)$ is the number of objects produced at $a_0$.
Note that the Heaviside step encodes that the objects do not exist before $a_0$.
We assume, consistent with Equation~(\ref{eqn:blueshift}), that each object produced will grow to the mass
\begin{align}
m(a, a_0) = m_0 \left(\frac{a}{a_0}\right)^k \label{eqn:mass_growth}
\end{align}
at time $a$, where
\begin{align}
  k &\equiv 3(1-\chi).
\end{align}
The normalized mass function $f(m_0, a_0)$ may be rewritten in terms of mass $m$ at time $a$ from Equation~(\ref{eqn:mass_growth})
\begin{align}
  m_0 &= \left(\frac{a_0}{a}\right)^k m  \\
  \d m_0 &= \left(\frac{a_0}{a}\right)^k~\d m.
\end{align}
The total number of objects at any time $a$ becomes 
\begin{align}
  N(a) = \int_0^\infty f\left[\left(\frac{a_0}{a}\right)^k m, a_0\right]~\left(\frac{a_0}{a}\right)^k N(a_0) \Theta_{a_0}(a)~\d m.%
\end{align}
We can now write a mass function for GEODEs at time $a$, given a birth spurt at $a_0$
\begin{align}
  \frac{\d N}{\d m} = f\left[\left(\frac{a_0}{a}\right)^k m, a_0\right]~\left(\frac{a_0}{a}\right)^k N(a_0)\Theta_{a_0}(a). \label{eqn:geode_mf_theta_spurt}
\end{align}

We have expressed our population in terms of the number of GEODEs produced at $a_0$.
We may re-express Equation~(\ref{eqn:geode_mf_theta_spurt}) in terms of the rate of production at $a_0$
\begin{align}
\frac{\d N}{\d m} = \int_0^a N(a_0')\delta(a_0-a_0') f\left[\left(\frac{a_0'}{a}\right)^k m, a_0'\right]~\left(\frac{a_0'}{a}\right)^k~\d a_0'. 
\end{align}
Our last step is to generalize the burst production rate $N(a_0')\delta(a_0-a_0')$ to a continuous production rate $R_\mathrm{prod}(a_0')$
\begin{align}
\frac{\d N}{\d m} = \int_0^a R_\mathrm{prod}(a_0) f_\mathrm{rem}\left[\left(\frac{a_0}{a}\right)^k m, a_0\right]~\left(\frac{a_0}{a}\right)^k~\d a_0. \label{eqn:inst_prod_model}
\end{align}
This is a general model for the apparent BHMF, when sourced by GEODEs.
We have dropped primes for clarity.
We emphasize that $f_\mathrm{rem}$ is the normalized distribution of collapse remnant objects, at the moment of formation $a_0'$.

\section{The apparent black hole mass function}
\label{sec:bhmf}
It is clear that any GEODE blueshift will powerfully impact the shape of the apparent black hole mass function (BHMF)
\begin{align}
  N_\mathrm{BH}(a) \equiv \int_0^\infty \frac{\d N_\mathrm{BH}}{\d M_\mathrm{BH}}~\d M_\mathrm{BH}, 
\end{align}
where $N_\mathrm{BH}(a)$ is the number count of BHs at $a$.
It follows from Equation~(\ref{eqn:geode-contribution}) that 
\begin{align}
\rho_s(a) = \frac{1}{\mathcal{V}}\int_0^\infty M_\mathrm{BH} \frac{\d N_\mathrm{BH}}{\d M_\mathrm{BH}}~\d M_\mathrm{BH}.
\label{eqn:local-interpretation}
\end{align}
In other words, the DE density is proportional to first moment of the BHMF, at fixed epoch.
In Appendix~\ref{sec:bhmf}, we develop a generic model of the BHMF, given a population of objects which evolve in mass from birth onward.
That model requires a remnant distribution at birth $f_\mathrm{rem}$ and a remnant production rate $R_\mathrm{prod}$.
In this section, we will make simplifying assumptions to arrive at tractable $f_\mathrm{rem}$ and $R_\mathrm{prod}$.  

To estimate $R_\mathrm{prod}$, we will use the mean mass density in stars.
The mean mass density of stars $\rho_*(a)$ from \citet{MadauDickinson2014} is a fit to data from $z \lesssim 8$.
In this redshift range, we expect the stellar density to track the initial GEODE number and distribution for two reasons.
First, since progenitor stars have cosmologically fixed mass, estimating progenitor production rate from the mean mass density makes sense.
Second, as shown by \citet{fryer2012compact}, the remnant distribution is determined primarily by the metalicity of the progenitor.
If we restrict our consideration to $z \lesssim 8$, the remnant distribution is essentially fixed in time relative to the progenitor distribution.

Let the average stellar mass at $a_0$ be
\begin{align}
\left<m_*(a_0)\right> = \int_{m_c}^\infty m_0 f_*(m_0, a_0)~\d m_0.
\end{align}
We may approximate the number of stars in a fiducial comoving volume $\mathcal{V}$ with $\rho_*(a_0)$ 
\begin{align}
   N_*(a_0) \simeq \rho_*(a_0)\frac{\mathcal{V}}{\left<m_*(a_0)\right>}.  \label{eqn:coarse_stellar_approx}
\end{align} 
Taking a derivative with respect to time gives
\begin{align}
  \frac{\d N_*}{\d a_0} = \frac{\d \rho_*}{\d a_0} \frac{\mathcal{V}}{\left<m_*(a_0)\right>} - \rho\frac{\mathcal{V}}{\left<m_*(a_0)\right>^2}\frac{\d \left<m_*(a_0)\right>}{\d a_0}. \label{eqn:distribution_via_madau} 
\end{align}
If we assume that the average stellar mass does not change much, then
\begin{align}
  \frac{\d N_*}{\d a_0} \simeq \frac{\d \rho_*}{\d a_0} \frac{\mathcal{V}}{\left<m_*(a_0)\right>}. \label{eqn:stellar_rate_approx}
\end{align}
This expression is our desired approximation for $R_\mathrm{prod}$. 
We now assume a time-independent Salpeter distribution $f_*$ for the stars
\begin{align}
  f_*(m_0, a_0) \equiv \frac{\alpha - 1}{m_c} \left(\frac{m_0}{m_c}\right)^{-\alpha} \Theta(m_0 - m_c), \label{eqn:salpeter}
\end{align}
where $m_c$ is the low mass cutoff.  This gives an expected mass
\begin{align}
  \left<m_*\right> = m_c\frac{\alpha - 1}{\alpha - 2}. \label{eqn:<m_0>}
\end{align}
Together, we find
\begin{align}
  R_\mathrm{prod}(a) \simeq \frac{\mathcal{V}(\alpha - 2)}{m_c(\alpha - 1)} \frac{\d \rho_*}{\d a}.
\end{align}

To approximate the normalized remnant distribution $f_\mathrm{rem}$, we again assume a Salpeter distribution, but with a distinct low mass cutoff $m_\ell$ and a high mass cap $m_h$
\begin{align}
  f_\mathrm{rem}(m_0, a_0) \equiv \frac{\alpha-1}{m_\ell^{1-\alpha} - m_h^{1-\alpha}} \Theta\left[(m - m_\ell)(m_h - m)\right] m^{-\alpha} \label{eqn:capped_remnants}.
\end{align}
Substitution of Equation~(\ref{eqn:salpeter}) into Equation~(\ref{eqn:inst_prod_model}), followed by substitution of Equation~(\ref{eqn:stellar_rate_approx}) for $R_\mathrm{prod}(a_0')$ gives
\begin{align}
\frac{\d n}{\d m}&=  \frac{m^{-\alpha} a^{k(\alpha - 1)} (\alpha-2)}{m_c\left(m_\ell^{1-\alpha} - m_h^{1-\alpha}\right)} \int_{a(m_\ell/m)^{1/k}}^{a_\mathrm{max}} \frac{\d \rho}{\d a_0} a_0^{k(1-\alpha)}~\d a_0,  \label{eqn:bhmf}
\end{align}
where we have defined
\begin{align}
  a_\mathrm{max} \equiv a\min\left[1, (m_h/m)^{1/k}\right].
\end{align}
Note that we have dropped primes and divided off the $\mathcal{V}$.
As anticipated, this expression is not correct in magnitude.
This can be seen in the $(k,m_h, m_\ell) \to (0,\infty, m_c)$ limit, where computing the average mass with Equation~(\ref{eqn:bhmf}) regenerates Equation~(\ref{eqn:coarse_stellar_approx}).
The expression is, however, the correct shape when used at times where the collapsed fraction of stellar mass is roughly constant.
Since the collapse fraction is insensitive to metalicity beyond $~10^{-1}Z_\odot$, our model should be reasonable for $z \leqslant 3$.

\section{Distribution function in redshift space}
\label{sec:redshift-distribution}
We wish to develop an estimate of the population of candidate merging compact objects.
To do this, we must model their formation, as distributed in time.
Unlike traditional compact objects, the observed GEODE population depends crucially on the formation times of the constituent GEODEs.
Thus, we must check systematic effects having to do with a more realistic rate of progenitor production with a the RW time-uniform assumption usually employed.

Consider the cosmic comoving star formation rate $\d \rho_*/\d t$, where $t$ is the standard RW time.
Since compact objects come from stellar progenitors, we will use this quantity to approximately track compact object formation.
If we assume that the average stellar mass is not changing too much over the times of interest, we have that
\begin{align}
  \frac{\d N_*}{\d t} \propto \frac{\d \rho_*}{\d t}.
\end{align}
We can express this quantity in redshift
\begin{align}
  \frac{\d N_*}{\d z} &= \frac{\d N_*}{\d t}\frac{\d t}{\d a}\frac{\d a}{\d z} \\
  &= -\frac{\d N_*}{\d t} \frac{1}{H(z)(1 + z)}.
\end{align}
We are interested in the number of objects formed between redshift $z$ and $z + \d z$.
The comoving volume of this shell in redshift is
\begin{align}
  \d V_\mathrm{C}(z) \propto 4\pi D_\mathrm{C}(z)^2 \frac{\d D_\mathrm{C}}{\d z}~\d z \qquad (\Omega_k = 0).
\end{align}
Here $D_\mathrm{C}(z)$ is the comoving distance from the present day to redshift $z$, as defined in \citet[][Equation~(15)]{hogg1999distance}.
Putting these together, we find that
\begin{align}
  \d N_* \propto \frac{\d \rho_*}{\d t} \frac{D_\mathrm{C}(z)^2}{H(z)(1 + z)} \frac{\d D_\mathrm{C}}{\d z}, \label{eqn:redshift-distribution}
\end{align}
which is our desired distribution.
Note that the time-uniform assumption is a simplification of Equation~(\ref{eqn:redshift-distribution}), with $\d\rho_*/\d t \equiv 1$.

\section{Approximate LIGO detector luminosity distance horizon with extreme mass ratios}
\label{sec:horizon}
\begin{figure}[t]
  \centering
  \includegraphics[width=\linewidth]{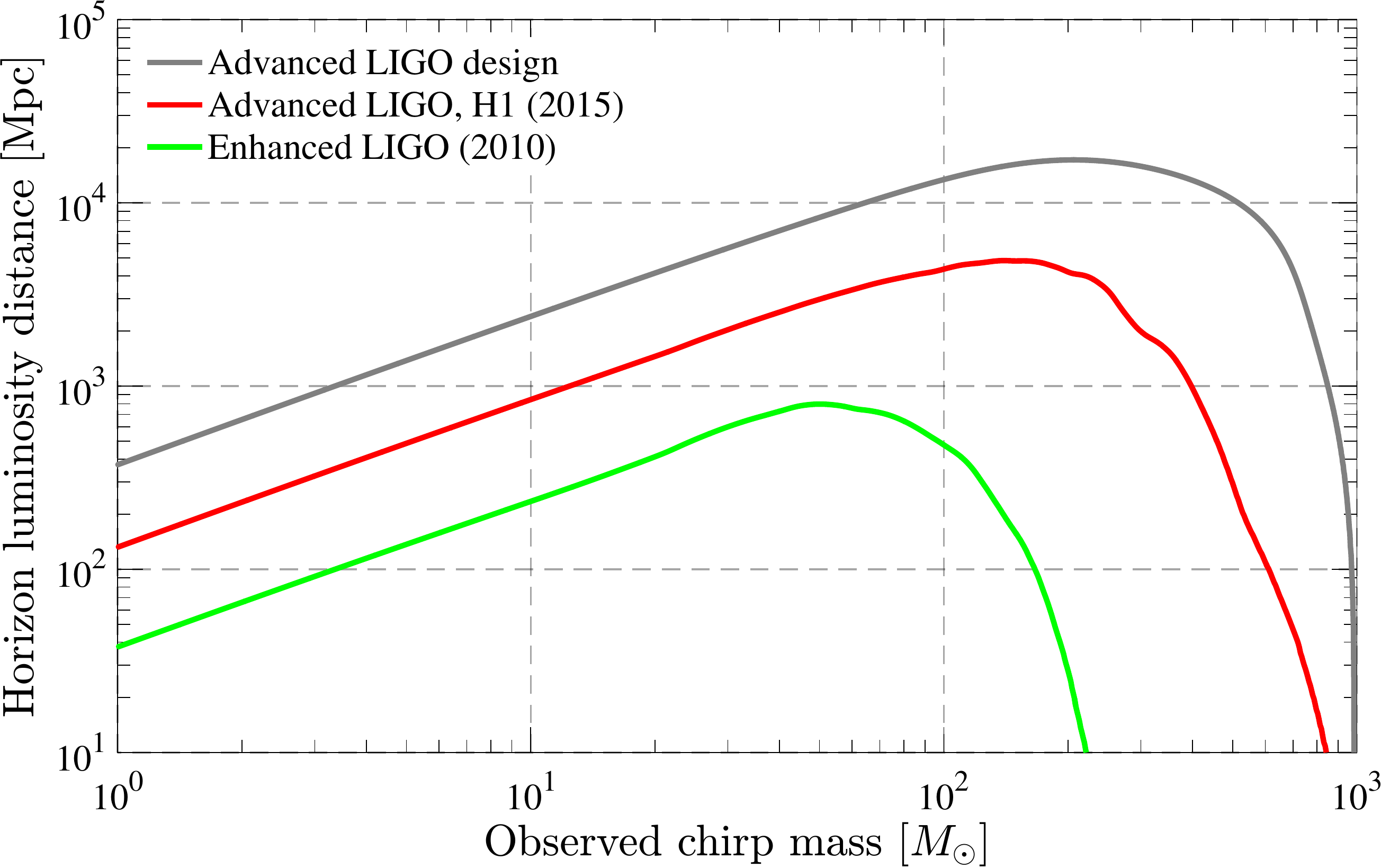}
  \caption{\label{fig:aligo_O1_sensitivity} Reproduction of \citet[][Figure 3]{martynov2016sensitivity}.  Sensitivity cutoff for coalescing compact binaries in aLIGO O1.  Assumes the most favorable detector orientation, head-on observation, equal chirp mass, and a signal to noise ratio of 8.}
\end{figure}%  
\begin{figure}[t]
  \centering
  \includegraphics[width=0.5\linewidth]{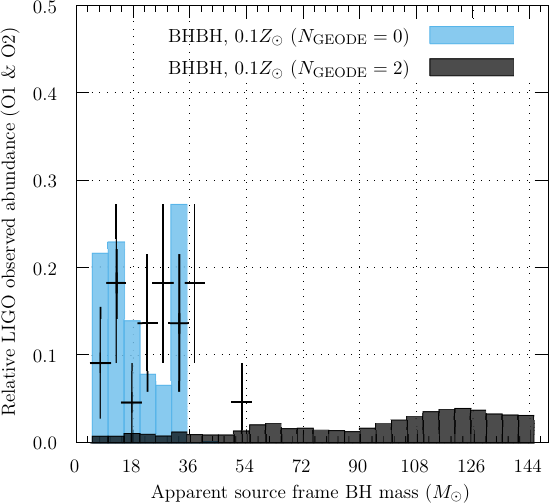}
  \caption{\label{fig:bhbh-hiz-geodegeode} Relative abundance vs. source frame mass for the low-metallicity $(Z=0.1Z_\odot)$ BH-BH initial conditions of \citet{de2015merger}, evolved to merger assuming that both remnant objects are GEODEs.
    GEODE population is shown in black (dark grey), control population evolved with cosmologically fixed mass is shown in sky blue (light grey).
    LIGO/Virgo GWTC-1 catalog overlaid with Poisson errors (black crosses).
    Populations have been selected according to LIGO O1 detector sensitivities following the procedure in Appendix~\ref{sec:horizon}.
    Graph is truncated at $50\%$ abundance for clarity, the control population bin extends to $100\%$.}
\end{figure}%
A coarse approximation of LIGO's detector sensitivity is given in \citet[][Figure 3]{martynov2016sensitivity} as horizon distance vs. chirp mass, for equal masses.
We reproduce their figure in our Figure~\ref{fig:aligo_O1_sensitivity}.
This horizon distance is a crude approximation to LIGO's actual sensitivities, which smoothly transition to zero around this given function in a complicated way.
We use this binary selection criterion to determine our estimates of LIGO observable populations of binaries.
Since a NS-GEODE mass ratio
\begin{align}
  q \equiv \frac{m_2}{m_1} \qquad (m_2 < m_1),
  \end{align}
can easily grow extremely small, in this Appendix, we coarsely extend this horizon estimate to mass ratio $q < 1$.

First, recall from \citet[][Equation~(9.166)]{schutz2009first} that the intrinsic bolometric luminosity of a binary inspiral in the linear regime is
\begin{align}
  \left<\frac{\d E}{\d t}\right> \propto (m_1 m_2)^2(m_1 + m_2) = m_1^5 q^2(1 + q)
\end{align}
where $m_1$ is the primary mass and $m_2 < m_1$ is the secondary mass.
Relative to an equal mass scenario, we have
\begin{align}
  \left<\frac{\d E}{\d t}\right>\Bigg|_{q < 1} = \frac{q^2(1 + q)}{2} \left<\frac{\d E}{\d t}\right>\Bigg|_{q = 1}.
\end{align}
Recall that the luminosity distance is defined as
\begin{align}
  D_\mathrm{L} \equiv \sqrt{\frac{L}{4\pi S}}
\end{align}
where $S$ is the bolometric flux and $L$ is the intrinsic bolometric luminosity.
Combining these two relations, we find that
\begin{align}
  D_\mathrm{L}\Big|_{q < 1} = \left(q \sqrt{\frac{1 + q}{2}}\right) D_\mathrm{L}\Big|_{q = 1}. \label{eqn:massratio-effect}
\end{align}
In other words, as the mass ratio becomes extreme, the radiated power drops approximately as $q^2$.
This causes the horizon distance to decrease by approximately $q$.

When we select whether LIGO can see a merger, we first compute the (asymmetric) chirp mass in the source frame.
We retrieve a luminosity distance horizon from \citet[][Figure 3]{martynov2016sensitivity} using the detector frame (redshifted) value of this chirp mass.
We then scale the luminosity distance horizon by $q(\sqrt{1 + q})/2$.
We convert this value to a redshift horizon assuming a Planck 2018~\citep{aghanim2018planck} median value cosmology, and use this to select visible mergers.

\section{Metallicity $0.1Z_\odot$ initial conditions}
\label{sec:lowZ}
\begin{figure}[t]
  \centering
  \includegraphics[width=0.5\linewidth]{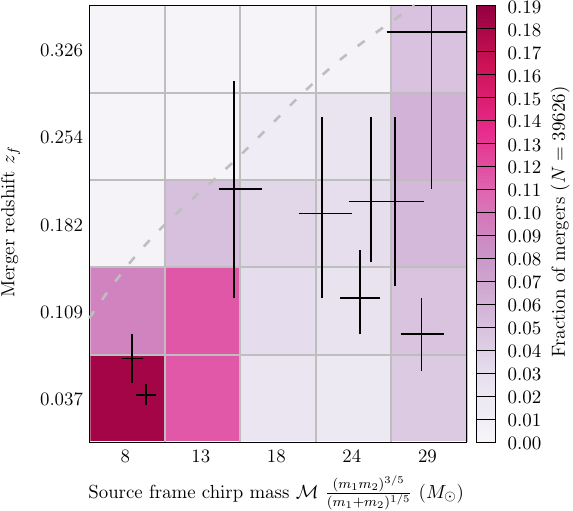}
  \caption{\label{fig:chirpz-bhbh-hiz-geodegeode} Merger redshift vs. source frame chirp mass for the low-metalicity $(Z=0.1Z_\odot)$ BH-BH initial conditions of \citet{de2015merger}, evolved without GEODEs $(N_\mathrm{GEODE}=0)$.
  }
\end{figure}%
In this section, we present low-metallicity initial conditions run as $N_\mathrm{GEODE} \in \{0,2\}$ systems.
Our purpose is to understand previous arguments (e.g. \citet{abbott2016astrophysical, belczynski2016first}) that low-metallicity populations could account for the observed LIGO masses with the CE formation channel.
Regeneration of this conclusion provides a useful systematic check on our study.

For consistency, we continue to assume a formation rate that tracks the SFR.
This will vastly overproduce $0.1Z_\odot$ objects, and so give very conservative bounds.
In Figure~\ref{fig:bhbh-hiz-geodegeode}, the control masses overlap well with LIGO observations.
Note that BHBH initial condition GEODE masses have little overlap.\footnote{NSNS $0.1Z_\odot$ initial condition GEODE masses are very similar to solar-metallicitiy GEODEs, so we do not display them}
In Figure~\ref{fig:chirpz-bhbh-hiz-geodegeode}, we consider the redshift-chirp distribution for the $0.1Z_\odot$ control $(N_\mathrm{GEODE}=0)$ population.
Note that there is continued overlap in redshift-chirp space.
This supports the notion that a low-metallicity progenitor population could produce intrinsically fixed-mass remnants sufficient to explain LIGO's observed events.

\clearpage
\bibliographystyle{aasjournal} \bibliography{thesis}

\end{document}